# Temperature dependence of the electronic structure of $Ca_3Cu_2O_4Cl_2$ Mott insulator

Haiwei Li[1,*], Shusen Ye[1,*], Jianfa Zhao[2,3,4], Changqing Jin[2,3,4], and Yayu Wang[1,5,†]

[1]*State Key Laboratory of Low Dimensional Quantum Physics, Department of Physics, Tsinghua University, Beijing 100084, P. R. China*

[2]*Beijing National Laboratory for Condensed Matter Physics, Institute of Physics, Chinese Academy of Sciences, Beijing 100190, P. R. China*

[3]*Songshan Lake Materials Laboratory, Dongguan 523808, P. R. China*

[4]*School of Physical Sciences, University of Chinese Academy of Sciences, Beijing 100049, P. R. China*

[5]*Frontier Science Center for Quantum Information, Beijing 100084, P. R. China*

[*]These authors contributed equally to this work.

[†]Email: yayuwang@tsinghua.edu.cn

**Abstract:** We use scanning tunneling microscopy to study the temperature evolution of electronic structure in $Ca_3Cu_2O_4Cl_2$ parent Mott insulator of cuprates. We find that the upper Hubbard band moves towards the Fermi energy with increasing temperature, while the charge transfer band remains basically unchanged. This leads to a reduction of the charge transfer gap size at high temperatures, and the rate of reduction is much faster than that of conventional semiconductors. Across the Neel temperature for antiferromagnetic order, there is no sudden change in the electronic structure. These results shed new light on the theoretical models about the parent Mott insulator of cuprates.

## 1 Introduction

The parent compound of the cuprate high-temperature superconductor is a prototypical Mott insulator [1-5], and the physical properties of its ground state are relatively well understood [6-11]. The copper $3d_{x^2-y^2}$ orbital on the $CuO_2$ plane is split into the upper and lower Hubbard bands by strong onsite Coulomb repulsion. Due to the hybridization with the oxygen $2p$ orbital, it is a charge transfer type Mott insulator, usually called a charge transfer insulator. The lowest energy excitation is from the hybridized charge transfer band (CTB) to the upper Hubbard band (UHB), and the energy difference between them is the charge transfer gap (CTG). Because of the superexchange between spins on neighboring Cu sites via the intermediary O ligand, the ground state of the Mott insulator has a long-range antiferromagnetic (AF) order. High temperature superconductivity emerges upon the suppression of AF order by charge carriers doped into the parent compound.

An important question that has yet to be thoroughly investigated is the electronic structure of cuprate Mott insulator at high temperatures. The size of the charge transfer gap is expected to be influenced by thermal activations in fundamentally different ways from conventional band insulators. It has been proposed theoretically that the short-range spin fluctuation renormalizes the charge sector and the activated spin excitation at finite temperature reduces the gap of electronic degree of freedom due to the convolution of charge and spin sectors [9]. Therefore, understanding the temperature evolution of the Mottness in cuprates will provide new insights for unravelling the mechanism of high-temperature superconductivity.

There have been a few experimental studies on the temperature dependence of the electronic structure of parent cuprate. Optical spectroscopy studied the evolution of the charge-transfer excitation spectrum with temperature in $La_2CuO_4$ [12] and observed the decrease of energy gap with temperature. Angle-resolved photoemission spectroscopy (ARPES) found strong temperature dependence of the quasiparticle peaks in the valence band in $Sr_2CuO_2Cl_2$ and $Ca_2CuO_2Cl_2$ (CCOC) [13]. Mid-infrared optical conductivity found a magnetic origin peak redshift and width broadening with increasing temperature [14]. Despite these investigations, some important issues about the Mott insulator remain unsolved, such as the temperature evolution of CTB and UHB positions, and whether the suppression of long-range AF order affects the electronic structure.

Scanning tunneling microscopy (STM) is an ideal technique to study how the electronic structure of cuprate parent Mott insulator evolves with temperature. The tunneling spectroscopy can detect both the occupied and unoccupied states, which allows an accurate measure of the charge transfer gap size. Compared with optical spectroscopy, STM can directly resolve the energy of the band edge relative to the Fermi energy ($E_F$), which can reveal the shift of the CTB and UHB with varied temperature. However, performing STM experiments on a Mott insulator over a wide temperature range is a highly challenging task due to the instability of the STM tip and the unavoidable thermal expansions.

In this paper, we use STM to study the temperature evolution of the electronic structure of $Ca_3Cu_2O_4Cl_2$, which is a double-layer counterpart of the more well-known CCOC parent Mott insulator of cuprates [15, 16]. We choose $Ca_3Cu_2O_4Cl_2$ in this work for two reasons. Firstly, its CTG size is significantly smaller than that of CCOC [17], so that the bias voltage involved in the experiment is smaller, making the STM tip more stable at high temperatures. Secondly, its Neel temperature for AF order $T_N$ = 230 K [18] is lower than $T_N$ = 250 K for CCOC [19], so it is easier to study the evolution of electronic structure across the $T_N$. We find that as the temperature increases, the UHB moves toward $E_F$ while the CTB remains nearly unchanged within the experimental uncertainty. As a consequence, the CTG size decreases with increasing temperature, and the reduction rate significantly exceeds that of conventional band insulators. When the temperature crosses the $T_N$, there is no abrupt change in the electronic structure.

## 2    Methods

The $Ca_3Cu_2O_4Cl_2$ crystal is grown by the self-flux method, as has been reported elsewhere [18]. The STM experiment is performed in an ultra-high vacuum chamber using an electrochemically etched tungsten tip. The tip has undergone a rigorous treatment and calibration process to ensure stability and reliability [20]. The crystal is cleaved at room temperature in the preparation chamber, and is then immediately transferred to the STM stage cooled by liquid nitrogen. The measurements start from the base temperature $T$ = 77 K, and higher temperatures are obtained by heating the STM stage. The STM topography is taken in the constant current mode with typical tunneling current $I_t$ = 10 pA, and the d$I$/d$V$ spectra are collected using a standard lock-in technique with modulation frequency $f$ = 447 Hz.

## 3    Experimental Results

The schematic crystal structure of $Ca_3Cu_2O_4Cl_2$ is shown in Fig. 1(a). It can be easily cleaved between the two weakly-bonded Cl atom layers, as shown by the gray plane. The inset of Fig. 1(b) displays the topography of a defect-free area, in which the regular square lattice of the exposed surface Cl atoms can be clearly resolved. The main panel of Fig. 1(b) shows a typical d$I$/d$V$ spectrum taken on the defect-free area at $T$ = 77 K, which exhibits a well-defined CTG. Figure 1(c) shows how to determine the band edge position from the spectrum in Fig. 2(b) by using a common practice [21]. We first vertically move the curve up by 2 times the maximum negative d$I$/d$V$ value, and then take the logarithm. Next, we calculate the average d$I$/d$V$ value and standard deviation within the CTG gap, and take linear fitting of the two band shoulders upon 4 standard deviations above the average value to avoid the influence of noise close to the band edge. The intersections of the linear fitting (red dashed lines) and the average value (blue dashed line) are used to extract the band edge of CTB and UHB.

Next, we explore the spatial variations of the d$I$/d$V$ spectra. We take about one hundred d$I$/d$V$ spectra at different locations sufficiently far from each other at 77 K and randomly display 20 spectra in Fig. 2(a). It can be seen that the spectral lineshapes are highly analogous to each other, but the positions of the two band edges fluctuate slightly in space due to the unavoidable defects that are several lattice constants away [22] or even in the underneath layers. To determine the size of the CTG, we extract the UHB and CTB band edge energies for each d$I$/d$V$ spectrum, as illustrated in Fig. 1(c), and make a statistical histogram in Fig. 2(b). It can be seen that the two band edge energies generally follow Gaussian distributions. The CTB lies in the range from -0.1 V to -0.4 V with an average value around -0.25 V, and the UHB lies between +1.3 V and +1.5 V with an average value around 1.4 V. The fluctuation of CTB is larger than that of UHB, which can also be seen directly from Fig. 2(a).

We use the averaged UHB and CTB energies in Fig. 2(b) to determine the CTG size, which is around 1.65 eV. This is larger than the 1.4 eV value obtained from the spatially averaged spectrum reported previously [17], in which the onset of the band edge is determined by the one that is closest to $E_F$. This can also be seen from Fig. 2(b), where the energy difference from the maximum CTB edge and minimum UHB edge is around 1.4 eV. Because the maximum CTB edge and minimum UHB edge are likely from different locations in space, the new method used here gives a more accurate estimate of the overall electronic structure.

We made the same measurements at 275 K, as shown in Fig. 2(c), and found that the electronic structure is similar to that at 77 K. The main difference between 275 K and 77 K is the position of the UHB band edge, which moves towards $E_F$ markedly. The statistical results in Fig. 2(d) also support this conclusion. The averaged CTB edge is almost the same as 77 K, but the UHB band edge reduces from 1.4 V to 1.05 V. There is no overlap between the distribution of the UHB band edge between 275 K and 77 K, proving that the change is due to the difference in temperature rather than spatial inhomogeneity.

To explore how the d$I$/d$V$ spectra evolve with temperature, we measured 5 temperature points from 77 K to 275 K, and the averaged spectra are displayed in Fig. 3(a). It reveals that as the temperature increases from 77 K to 275 K, the UHB edge position drops significantly, whereas the CTB only has a slight movement, which is consistent with the ARPES result [13]. Figure 3(b) summarizes the average value and the variation of CTB and UHB edge positions with temperature. It is worth noting that across the $T_N$ = 230 K of $Ca_3Cu_2O_4Cl_2$ [18], there is no abrupt change in the electronic structure. Instead, the overall evolution is very smooth over the whole temperature range.

In Fig. 4(a), we compare the temperature dependence of the CTG gap size of $Ca_3Cu_2O_4Cl_2$ with the band gap sizes of conventional semiconductor Si and GaAs. Because their gap sizes are different, we normalize the gap values to that taken at 77 K for each material. It can be seen that although the energy gaps of all three materials decrease with increasing temperature, the gap size of $Ca_3Cu_2O_4Cl_2$ decreases nearly 4 times faster than that of Si and GaAs. The decrease of gap size in conventional semiconductor is due to the thermal expansion of lattice and the electron-phonon coupling effect [23]. In Mott insulators, on the other hand, the excited state at high temperatures are determined by novel strong correlation mechanisms, such as spin fluctuations [9].

## 4  Discussion and conclusions

The STM results shown above provide new clues about the electronic structure evolution of parent cuprate with temperature. Below we summarize the main experimental findings and their implications to theoretical models about the cuprate Mott insulator.

The first finding is that the CTG size drops with increasing temperature at a much faster rate than that in conventional semiconductors. Theoretically, the 2D Hubbard model uses the slave-fermion method to calculate the Mott gap size with varied temperature [8, 9]. This method predicts that as the temperature rises, the fluctuation of the spin sector in the Mott insulator increases, resulting in a broadening of the energy spectrum. As a consequence, the Mott gap size reduces, as shown in Fig. 4(b). Figure 4(a) compares our experimental data on $Ca_3Cu_2O_4Cl_2$ with the calculated Mott gap of the 2D Hubbard model with a hopping integral $t$ = 0.35 eV [24]. The theoretically calculated gap reduction is faster than that of conventional semiconductors, but is still slightly slower than our experimental data. This discrepancy could be due to the difference between the *t* and *U* values used in the theoretical calculations and the real values in $Ca_3Cu_2O_4Cl_2$.

The second finding is that only the UHB moves to lower energy with increasing temperature, whereas the CTB is almost unchanged. This may be due to the different origins of the two bands. The UHB is composed of Cu 3*d* orbital, while the CTB is mainly composed of O 2*p* orbitals in the form of Zhang-Rice singlet [7]. Because the local spin is contributed by Cu 3*d* electrons, the influence of temperature on the spin fluctuation is mainly reflected by the UHB. In contrast, although the O 2*p* electrons can interact with the AF background through Zhang-Rice singlet, such indirect interaction is much less affected by temperature. Therefore, the spin fluctuation induced band broadening mainly happen at the UHB and LHB but not the CTB, as shown in Fig. 4(b).

The third finding is across the $T_N$, there is no abrupt change of electronic structure. It proves that the parent compound of cuprate belongs to the Mott insulator rather than the Slater insulator [25, 26] that is tied with the long-range AF order. Theoretically, the long-range AF order is likely to affect the spin fluctuations, and thus the size of the CTG. However, both experiment [27] and theory [28] show that the antiferromagnetism caused by strong correlation effect is different from the usual AF order, and our results clearly show that the antiferromagnetism in $Ca_3Cu_2O_4Cl_2$ has no obvious effect on the electronic structure. This implies that the electronic structure of parent cuprate can be described by the 2D Hubbard model, in which long-range AF order is absent at finite temperature according to the Mermin–Wagner theorem [29].

In summary, our STM data reveals how the electronic structure of parent cuprate evolves with temperature. Both the evolution of the UHB and CTB positions, as well as the absence of abrupt change across the $T_N$, can be explained at the qualitatively level by the spin fluctuations in Mott insulator [9]. The quantitative results of the CTG size and its dependence on temperature provide new experimental basis for the theoretical models for Mott insulators.

**Acknowledgements**

We acknowledge helpful discussions with Xingjie Han and Tao Xiang. This work was supported by the MOST of China grant No. 2017YFA0302900 and the Basic Science Center Project of NSFC under grant No. 51788104. It is supported in part by the Beijing Advanced Innovation Center for Future Chip (ICFC).

**Figure Captions**

FIG. 1. (a) Schematic crystal structure of $Ca_3Cu_2O_4Cl_2$. The crystal can be easily cleaved between two adjacent Cl layers, as indicated by the grey planes. (b) A typical d$I$/d$V$ spectrum taken on the cleaved surface of the $Ca_3Cu_2O_4Cl_2$ at 77 K. (Inset) Topographic image (50 Å × 50 Å) taken at $V_b$ = -2.5 V and $I_t$ = 10 pA. (c) Determination of the CTB and UHB band edges for the spectrum in (b).

FIG. 2. (a) Twenty d$I$/d$V$ spectra taken at different positions sufficiently far from each other at 77 K. (b) Statistics of the energy of the band edge of the CTB and UHB at 77 K. (c) Twenty d$I$/d$V$ spectra taken at different positions at 275 K. (d) Statistics of the energy of the band edge of the CTB and UHB at 275 K.

FIG. 3. (a) The logarithm plot of the spatial averaged d$I$/d$V$ curves at 77 K, 125 K, 175 K, 225 K and 275 K. (b) Evolution of the band edge of the CTB and UHB with temperature. The triangle data points represent the statistical average value of the band edge at each temperature point, and the error bar represents the range of a standard deviation.

FIG. 4. (a) Comparison of the temperature dependence of the gap sizes in $Ca_3Cu_2O_4Cl_2$, GaAs, Si, and theory prediction using the slave-fermion method. (b) Schematic band structure of parent cuprate Mott insulator at high temperature. The UHB and LHB are broadened due to spin fluctuation, but the CTB is nearly unaffected. The reduction of the CTG size is mainly due to the UHB broadening.

**Author Contributions**

Haiwei Li and Shusen Ye carried out the STM experiments. Jianfa Zhao and Changqing Jin grew the $Ca_3Cu_2O_4Cl_2$ single crystals. Yayu Wang designed the project and prepared the manuscript. All authors have read and approved the final version of the manuscript.

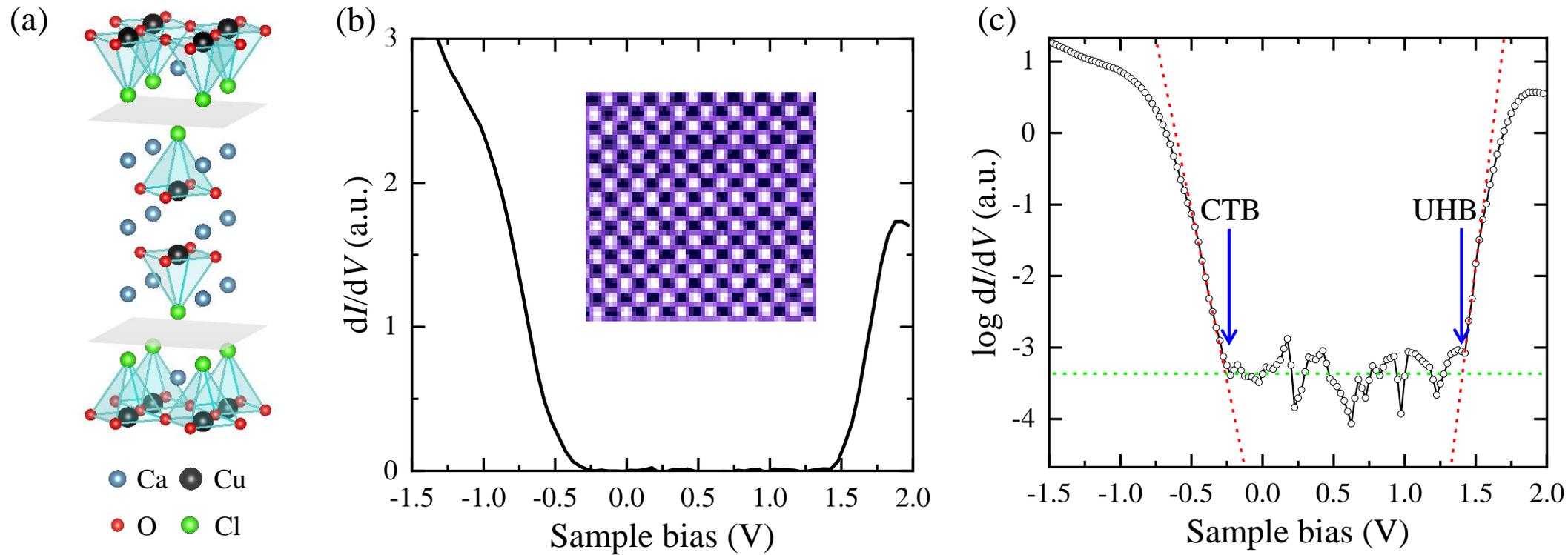

Figure 1

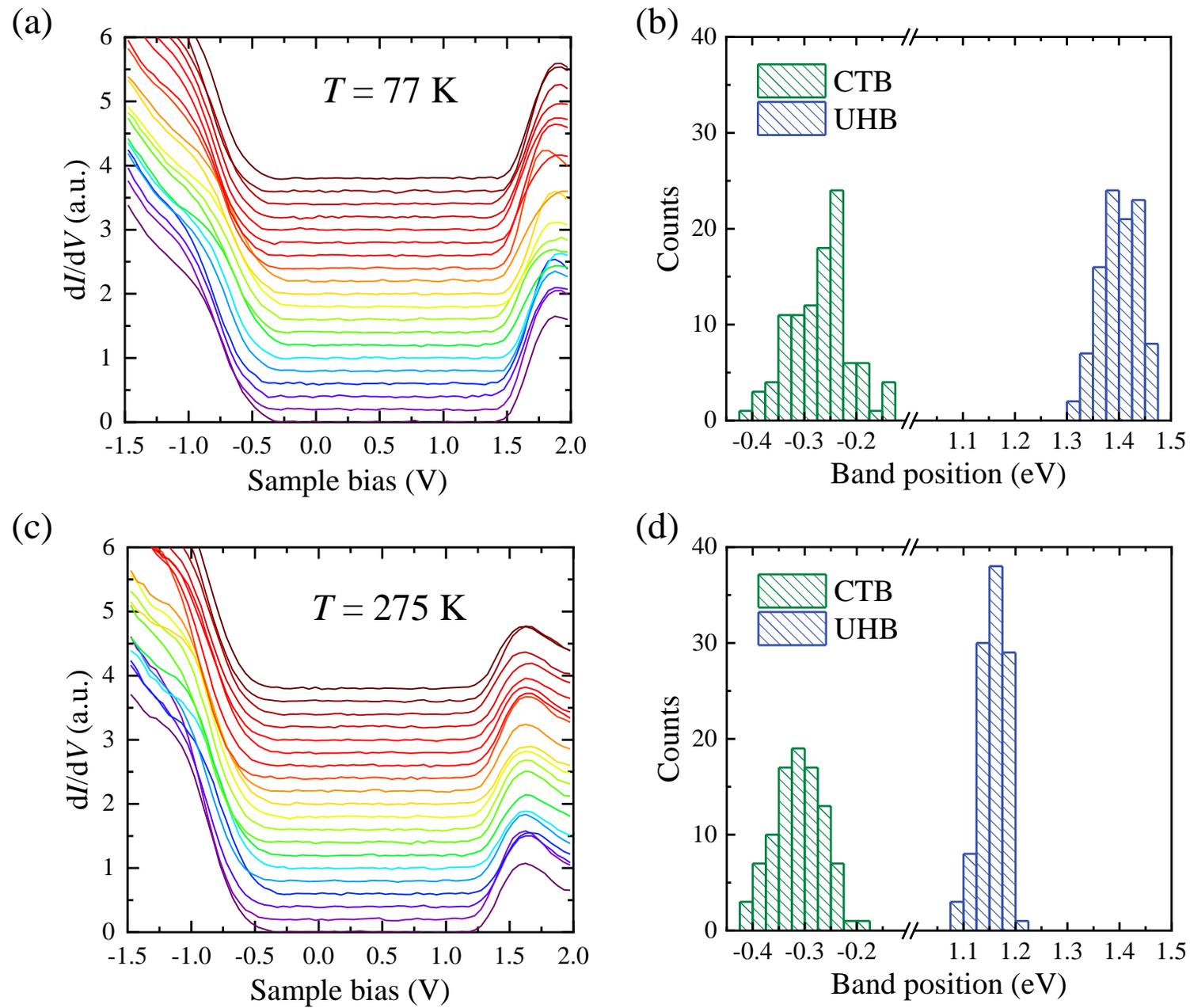

Figure 2

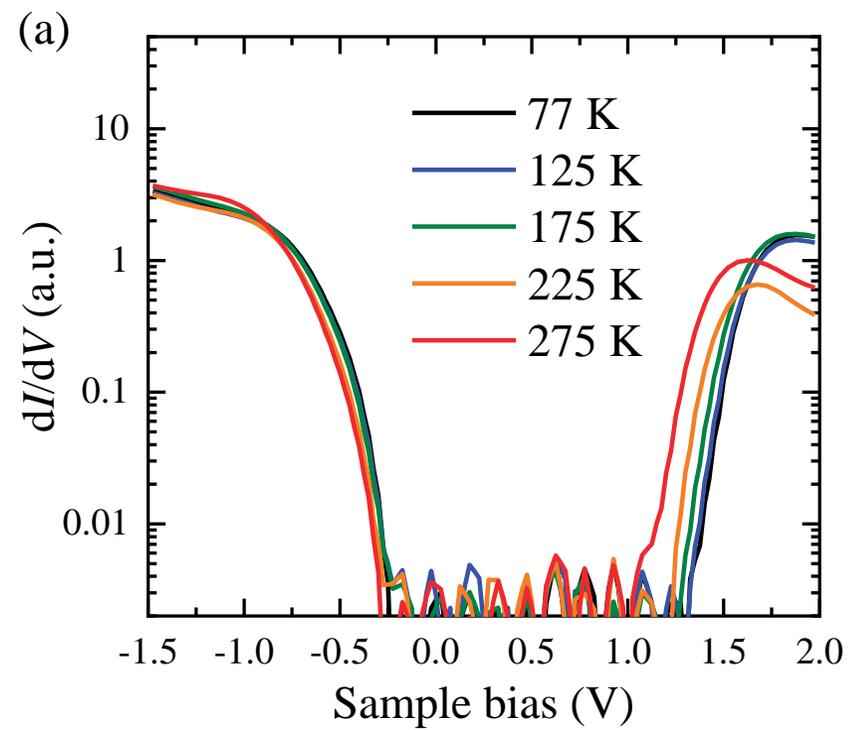 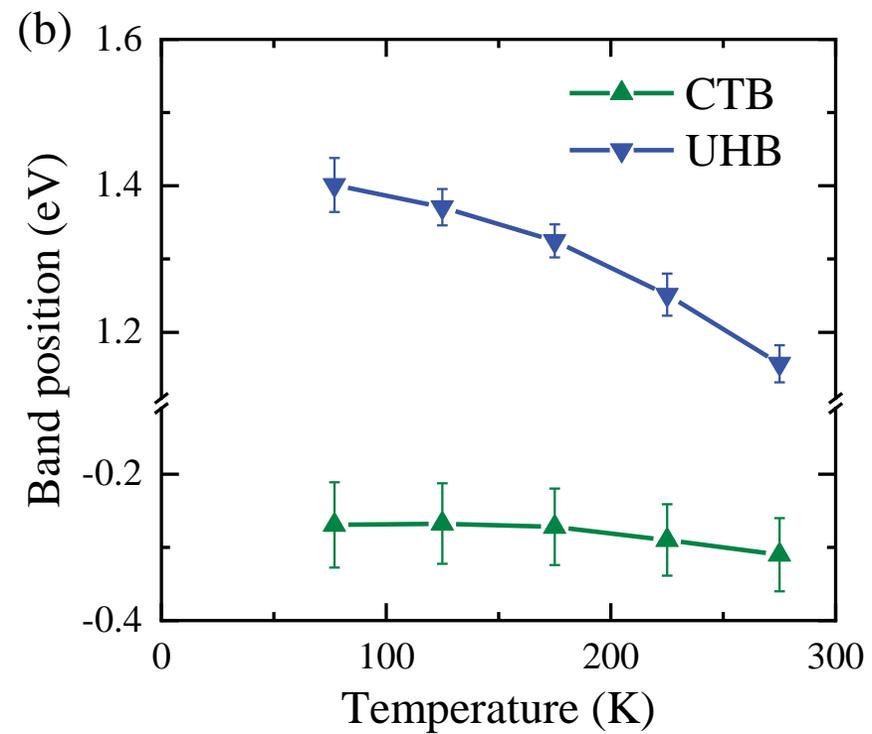

Figure 3

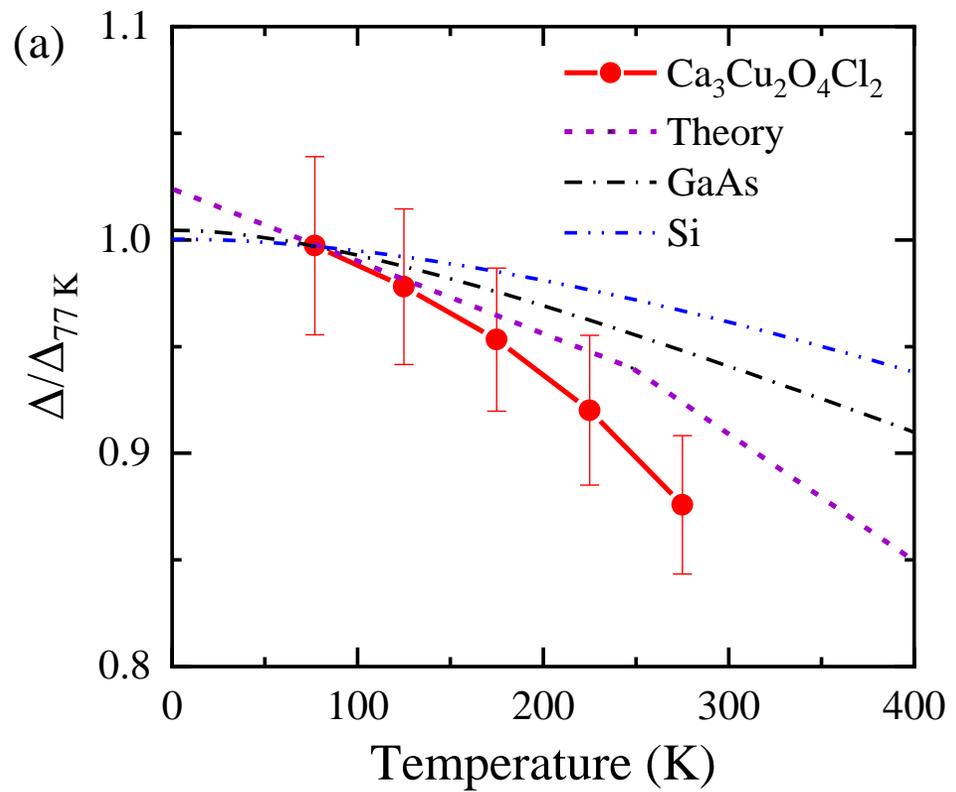 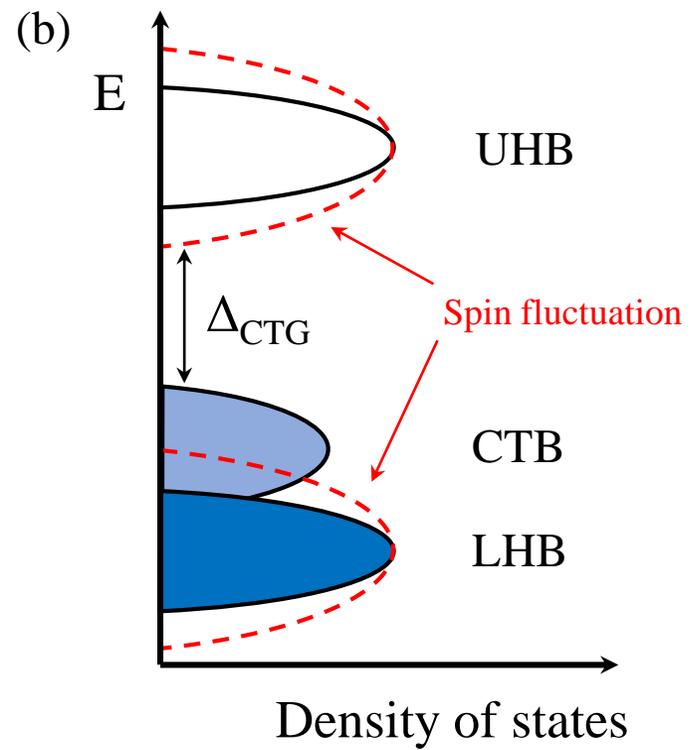

Figure 4